\documentclass{ws-ijmpb}

\usepackage{graphicx}
\usepackage{amssymb}

\begin{document}

\markboth{A.~Sherman \& M.~Schreiber} {Incommensurate spin dynamics in
underdoped cuprate perovskites}

\title{INCOMMENSURATE SPIN DYNAMICS\\ IN UNDERDOPED CUPRATE PEROVSKITES}

\author{A.~Sherman}

\address{Institute of Physics, University of Tartu, Riia 142, 51014
Tartu, Estonia\\alexei@fi.tartu.ee}

\author{M.~Schreiber}

\address{Institut f\"ur Physik, Technische Universit\"at, D-09107
Chemnitz, Federal Republic of Germany}

\maketitle

\begin{history}
\received{Day Month Year} \revised{Day Month Year}
%\accepted{(Day Month Year)}
%\comby{(xxxxxxxxxx)}
\end{history}

\begin{abstract}
The incommensurate magnetic response observed in normal-state cuprate
pe\-rov\-ski\-tes is interpreted based on the projection operator
formalism and the $t$-$J$ model of Cu-O planes. In agreement with
experiment the calculated dispersion of maxima in the susceptibility
has the shape of two parabolas with upward and downward branches which
converge at the antiferromagnetic wave vector. The maxima are located
at the momenta $(\frac{1}{2},\frac{1}{2}\pm\delta)$,
$(\frac{1}{2}\pm\delta,\frac{1}{2})$ and at
$(\frac{1}{2}\pm\delta,\frac{1}{2}\pm\delta)$,
$(\frac{1}{2}\pm\delta,\frac{1}{2}\mp\delta)$ in the lower and upper
parabolas, respectively. The upper parabola reflects the dispersion of
magnetic excitations of the localized Cu spins, while the lower
parabola arises due to a dip in the spin-excitation damping at the
antiferromagnetic wave vector. For moderate doping this dip stems from
the weakness of the interaction between the spin excitations and holes
near the hot spots. The frequency dependence of the susceptibility is
shown to depend strongly on the hole bandwidth and damping and varies
from the shape observed in YBa$_2$Cu$_3$O$_{7-y}$ to that inherent in
La$_{2-x}$Sr$_x$CuO$_4$.
\end{abstract}


\section{Introduction}
One of the most interesting features of the inelastic neutron
scattering in lanthanum cuprates is that for hole concentrations $x
\gtrsim 0.04$, low temperatures and small energy transfers the
scattering is peaked at incommensurate momenta
$(\frac{1}{2},\frac{1}{2}\pm\delta)$,
$(\frac{1}{2}\pm\delta,\frac{1}{2})$ in the reciprocal lattice units
$2\pi/a$ with the lattice period $a$.\cite{Yoshizawa} For $x\lesssim
0.12$ the incommensurability parameter $\delta$ is approximately equal
to $x$.\cite{Yamada} For larger $x$ the parameter saturates near the
value $\delta\approx 0.12$. The incommensurate response was observed
both below and above $T_c$.\cite{Mason93} Recently the analogous
low-frequency incommensurability was observed also in
YBa$_2$Cu$_3$O$_{7-y}$.\cite{Dai} This gives ground to suppose that the
incommensurability is a common feature of cuprate perovskites which
does not depend on subtle details of the energy structure. However, for
larger frequencies the susceptibility differs essentially in these two
types of cuprates. In the underdoped YBa$_2$Cu$_3$O$_{7-y}$ and some
other cu\-p\-ra\-tes both below and above $T_c$ a pronounced maximum is
observed at frequencies $\omega_r=25-40$~meV.\cite{Bourges} In the
momentum space the magnetic response is sharply peaked at the
antiferromagnetic wave vector ${\bf Q}=(\frac{1}{2},\frac{1}{2})$ for
this frequency. Contrastingly, no maximum at $\omega_r$ was observed in
lanthanum cuprates. Instead for low temperatures and frequencies of
several millielectronvolts a broad feature was detected.\cite{Aeppli}
For even larger frequencies the magnetic response becomes again
incommensurate in both types of cuprates with peaks located at
$(\frac{1}{2}\pm\delta,\frac{1}{2}\pm\delta)$,
$(\frac{1}{2}\pm\delta,\frac{1}{2}\mp
\delta)$.\cite{Bourges,Hayden96,Hayden04,Tranquada} In contrast to the
low-frequency incommensurability in which the incommensurability
parameter decreases with increasing frequency, the parameter of the
high-frequency incommensurability grows or remains practically
unchanged with frequency. Thus, the dispersion of maxima in the
susceptibility resembles two parabolas with upward- and
downward-directed branches which converge at {\bf Q} and near the
frequency $\omega_r$.\cite{Dai,Tranquada}

The nature of the magnetic incommensurability is the subject of active
discussion now. The most frequently used approaches for its explanation
are based on the picture of itinerant electrons with the susceptibility
calculated in the random phase approximation\cite{Liu,Brinckmann} and
on the stripe domain picture.\cite{Tranquada,Hizhnyakov} In the former
approach the low-frequency incommensurability is connected with the
Fermi surface nesting in the normal state or with the nesting in
constant-energy contours in the superconducting case. This imposes
rather stringent requirements on the electron energy spectrum, since
the nesting has to persist in the range of hole concentrations $0.04
\lesssim x\lesssim 0.18$ where the incommensurability is observed and
the nesting momentum has to change in a specific manner with doping to
ensure the known dependence of the incommensurability parameter
$\delta$ on $x$. It is unlikely that these conditions are fulfilled in
La$_{2-x}$Sr$_x$CuO$_4$.\cite{Ino} Besides, the applicability of the
picture of itinerant electrons for underdoped cuprates casts doubts. As
for the second notion, it should be noted that in the elastic neutron
scattering the charge-density wave connected with stripes is observed
only in crystals with the low-temperature tetragonal or the
low-temperature less-orthorhombic phases (La$_{2-x}$Ba$_x$CuO$_4$
\linebreak[4] and La$_{2-y-x}$Nd$_y$Sr$_x$CuO$_4$) and is not observed
in the crystal La$_{2-x}$Sr$_x$CuO$_4$ in the low-temperature
orthorhombic phase.\cite{Kimura} At the same time the magnetic
incommensurability is similar in these phases. It can be supposed that
the magnetic incommensurability is the cause rather than the effect of
stripes which are formed with an assistance of phonons.

In the present work the general formula for the magnetic susceptibility
derived in the projection operator formalism\cite{Mori} is used. For
the description of spin excitations in the doped antiferromagnet the
$t$-$J$ model of a Cu-O plane is employed. In this approach the
mentioned peculiarities of the magnetic properties of cuprates are
reproduced including the proper frequency and momentum location of the
susceptibility maxima. The incommensurability for $\omega>\omega_r$ is
connected with the dispersion of spin
excitations.\cite{Barzykin,Sherman02} The incommensurability for lower
frequencies is related to the dip in the spin-excitation damping at
${\bf Q}$. For small $x$ the dip appears due to the nesting of the hole
pockets around $(\pm\frac{1}{4},\pm\frac{1}{4})$ forming the Fermi
surface.\cite{Sherman04} For moderate doping this dip stems from the
weakness of the interaction between the spin excitations and holes near
the hot spots -- the intersection points of the Fermi surface and the
boundary of the magnetic Brillouin zone. Such a weak interaction
follows from the fact that due to a short-range interaction between
holes and spins a decaying site spin excitation creates a fermion pair
with components residing on the same and neighbor sites. The
spin-excitation damping was found to depend strongly on details of the
hole dispersion, bandwidth and damping, so that the change in these
characteristics leads to the conversion of well-defined spin
excitations to overdamped ones. As this takes place, the frequency
dependence of the susceptibility at {\bf Q} is transformed from a
pronounced maximum\cite{Bourges} at $\omega_r$ which is inherent in
underdoped YBa$_2$Cu$_3$O$_{7-y}$ to a broad low-frequency feature
characteristic for lanthanum cuprates.\cite{Aeppli} The increased
spin-excitation damping has no marked effect on the low-frequency
incommensurability, however for $\omega>\omega_r$ the incommensurate
peaks are shifted to {\bf Q} and form a broad maximum. Such form of the
momentum dependence of the susceptibility is also observed
experimentally.\cite{Endoh}

\section{Main formulas}
The imaginary part of the magnetic susceptibility which determines the
cross-section of the magnetic scattering\cite{Kastner} is calculated
from the relations $\chi''({\bf k}\omega)=-4\mu_B^2\Im\langle\langle
s^z_{\bf k}|s^z_{\bf -k}\rangle\rangle_\omega$, $\langle\langle
s^z_{\bf k}|s^z_{\bf -k}\rangle\rangle_\omega=\omega((s^z_{\bf
k}|s^z_{\bf -k}))_\omega-(s^z_{\bf k},s^z_{\bf -k})$. Here $\mu_B$ is
the Bohr magneton, $\langle\langle s^z_{\bf k}|s^z_{\bf
-k}\rangle\rangle_\omega$ and $((s^z_{\bf k}|s^z_{\bf -k}))_\omega$ are
the Fourier transforms of the retarded Green's and Kubo's relaxation
functions, $$\langle\langle s^z_{\bf k}|s^z_{\bf
-k}\rangle\rangle_t=-i\theta(t)\langle[s^z_{\bf k}(t),s^z_{\bf
-k}]\rangle,\quad ((s^z_{\bf k}|s^z_{\bf -k}))_t=\theta(t)\int_t^\infty
dt'\langle[s^z_{\bf k}(t'),s^z_{\bf -k}]\rangle,$$ $s^z_{\bf
k}=N^{-1/2}\sum_{\bf n}e^{-i{\bf kn}}s^z_{\bf n}$ with the number of
sites $N$ and the $z$ component of the spin $s^z_{\bf n}$ on the
lattice site {\bf n}, for arbitrary operators $A$ and $B$
$(A,B)=i\int_0^\infty dt\langle[A(t),B]\rangle$ where the angular
brackets denote the statistical averaging and $A(t)=e^{iHt}Ae^{-iHt}$
with the Hamiltonian $H$.

Using the projection operator technique\cite{Mori} the relaxation
function $((s^z_{\bf k}|s^z_{\bf -k}))_\omega$ can be calculated from
the recursive relations
\begin{equation}\label{cf}
R_n(\omega)=[\omega-E_n-F_nR_{n+1}(\omega)]^{-1}, \quad n=0,1,2,\ldots
\end{equation}
where $R_n(\omega)$ is the Laplace transform of
$R_n(t)=(A_{nt},A_n^\dagger)(A_{n},A_n^\dagger)^{-1}$, the time
dependence in $A_{nt}$ is determined by the relation
$$i\frac{d}{dt}A_{nt}=\prod_{k=0}^{n-1}(1-P_k)[A_{nt},H], \quad
A_{n,t=0}=A_n$$ with the projection operators $P_k$ defined as
$P_kB=(B,A_k^\dagger) (A_k,A_k^\dagger)^{-1}A_k$. The parameters $E_n$
and $F_n$ in relations (\ref{cf}) and operators $A_n$ in the functions
$R_n(t)$ are calculated recursively using the procedure\cite{Sherman02}
\begin{eqnarray}
&&[A_n,H]=E_nA_n+A_{n+1}+F_{n-1}A_{n-1},\quad
E_n=([A_n,H],A_n^\dagger)(A_n,A_n^\dagger)^{-1}, \nonumber\\
&&\label{Lanczos}\\
&&F_{n-1}=(A_n,A_n^\dagger)(A_{n-1},A_{n-1}^\dagger)^{-1},\quad
F_{-1}=0. \nonumber
\end{eqnarray}
As the starting operator for this procedure we set $A_0=s^z_{\bf k}$.
In this case $((s^z_{\bf k}|s^z_{\bf -k}))_\omega=(s^z_{\bf k},s^z_{\bf
-k})R_0(\omega)$ where $R_0(\omega)$ is calculated from Eq.~(\ref{cf}).

To describe the spin excitations of Cu-O planes which determine the
magnetic properties of cuprates\cite{Kastner} the $t$-$J$
model\cite{Izyumov} is used. The model was shown to describe correctly
the low-energy part of the spectrum of the realistic extended Hubbard
model.\cite{Zhang,Jefferson} The Hamiltonian of the two-dimensional
$t$-$J$ model reads
\begin{equation}\label{hamiltonian}
H=\sum_{\bf nm\sigma}t_{\bf nm}a^\dagger_{\bf n\sigma}a_{\bf
m\sigma}+\frac{1}{2}\sum_{\bf nm}J_{\bf nm}{\bf s_n s_m},
\end{equation}
where $a_{\bf n\sigma}=|{\bf n}\sigma\rangle\langle{\bf n}0|$ is the
hole annihilation operator, {\bf n} and {\bf m} label sites of the
square lattice, $\sigma=\pm 1$ is the spin projection, $J_{\bf nm}$ and
$t_{\bf nm}$ are the exchange and hopping constants, respectively,
$|{\bf n}\sigma\rangle$ and $|{\bf n}0\rangle$ are site states
corresponding to the absence and presence of a hole on the site. These
states may be considered as linear combinations of the products of the
$3d_{x^2-y^2}$ copper and $2p_\sigma$ oxygen orbitals of the extended
Hubbard model.\cite{Jefferson} The spin-$\frac{1}{2}$ operators can be
written as $s^z_{\bf n}=\frac{1}{2}\sum_\sigma\sigma|{\bf
n}\sigma\rangle\langle{\bf n}\sigma|$ and $s^\sigma_{\bf n}=|{\bf
n}\sigma\rangle\langle{\bf n},-\sigma|$.

With Hamiltonian (\ref{hamiltonian}) and $A_0=s^z_{\bf k}$ we find from
Eq.~(\ref{Lanczos})
\begin{eqnarray}
&&E_0(s^z_{\bf k},s^z_{\bf -k})=(i\dot{s}^z_{\bf k},s^z_{\bf
-k})=\langle[s^z_{\bf k},s^z_{\bf -k}]\rangle=0, \nonumber\\
&&A_1=A_1^s+A_1^h=\frac{1}{2\sqrt{N}}\sum_{\bf l}e^{-i{\bf
kl}}\bigg[\sum_{\bf nm}J_{\bf mn}(\delta_{\bf ln}-\delta_{\bf
lm})s_{\bf n}^{+1}s_{\bf m}^{-1}\label{fstep}\\
&&\hspace{1.5em}+\sum_{\bf nm\sigma}t_{\bf mn}(\delta_{\bf
lm}-\delta_{\bf ln})\sigma a_{\bf n\sigma}^\dagger a_{\bf
m\sigma}\bigg],\nonumber
\end{eqnarray}
where $i\dot{s}^z_{\bf k}=[s^z_{\bf k},H]$. To obtain a tractable form
for the spin-excitation damping it is convenient to approximate the
quantity $(A_{1t},A_1^\dagger)$ in the $R_1(\omega)$ by the sum
$$(A_1^h(t),A_1^{h\dagger})+(A_{1t}^s,A_1^{s\dagger})$$ where the first
term describes the influence of holes on the spin excitations.
Continuing calculations (\ref{Lanczos}) with the second term of the sum
we get
\begin{equation}\label{sstep}
F_0=4JC_1(\gamma_{\bf k}-1)(s^z_{\bf k},s^z_{\bf -k})^{-1}, \quad
E_1=0,
\end{equation}
where only the nearest neighbor interaction between spins was taken
into account, $J_{\bf nm}=J\sum_{\bf a}\delta_{\bf n,m+a}$, the four
vectors {\bf a} connect the nearest neighbor sites, $C_1=\langle
s^{+1}_{\bf n}s^{-1}_{\bf n+a}\rangle$ is the spin correlation on
neighbor sites and $\gamma_{\bf k}=\frac{1}{2}[\cos(k_x)+\cos(k_y)]$.

To calculate the quantity $(s^z_{\bf k},s^z_{\bf -k})$ let us notice
that in accord with procedure (\ref{Lanczos}) the interruption of
calculations at this stage actually means that $(A_2,A_2^\dagger)$ in
the parameter $F_1$ is set to zero. Here $A_2=i^2\ddot{s}^z_{\bf
k}-F_0s^z_{\bf k}$. The substitution of this expression into
$(A_2,A_2^\dagger)=0$ gives an equation for $(s^z_{\bf k},s^z_{\bf
-k})$. Using the decoupling in calculating $i^2\ddot{s}^z_{\bf k}$ we
get\cite{Sherman02}
\begin{equation}\label{inpr}
(s^z_{\bf k},s^z_{\bf -k})^{-1}=4\alpha J(\Delta+1+\gamma_{\bf k}),
\end{equation}
where $\alpha\sim 1$ is the decoupling parameter.\cite{Kondo} The
meaning of the parameter $\Delta$, which can be expressed in terms of
spin correlations, will be discussed later.

Using the decoupling in $(A_1^h(t),A_1^{h\dagger})$ we find from the
above formulas
\begin{equation}\label{chi}
\chi''({\bf k}\omega)=-\frac{4\mu^2_B\omega\Im R({\bf k}\omega)
}{[\omega^2-\omega f_{\bf k}\Re R({\bf k}\omega)-\omega^2_{\bf
k}]^2+[\omega f_{\bf k}\Im R({\bf k}\omega)]^2},
\end{equation}
where
\begin{eqnarray}
&&f_{\bf k}^{-1}=4J|C_1|(1-\gamma_{\bf k}), \quad \omega_{\bf
k}^2=16J^2\alpha|C_1|(1-\gamma_{\bf k})(\Delta+1+
 \gamma_{\bf k}), \nonumber\\
&&\Im R({\bf k}\omega)=\frac{8\pi\omega^2_{\bf k}}{N}
 \sum_{\bf k'}g_{\bf kk'}^2\int_{-\infty}^\infty d\omega' A({\bf
 k'}\omega') \label{chitJ}\\
&&\hspace{4em}\times A({\bf k+k'},\omega+\omega')
 \frac{n_F(\omega+\omega')-n_F(\omega')}{\omega},\nonumber
\end{eqnarray}
the interaction constant $g_{\bf kk'}=t_{\bf k'}-t_{\bf k+k'}$ with
$t_{\bf k}=\sum_{\bf n}e^{i{\bf k(n-m)}}t_{\bf nm}$,
$n_F(\omega)=[\exp(\omega/T)+1]^{-1}$, $T$ is the temperature and
$A({\bf k}\omega)$ is the hole spectral function. Since the incoherent
part of the spectral function is unlikely to lead to sharp structure in
$\chi''$, only the coherent part of $A({\bf k}\omega)$ is taken into
account in this work,
\begin{equation}\label{hsf}
A({\bf k}\omega)=\frac{\eta/\pi}{(\omega-\varepsilon_{\bf
k}+\mu)^2+\eta^2}.
\end{equation}
Here $\mu$ is the chemical potential, $\eta$ is the artificial
broadening, and $\varepsilon_{\bf k}$ is the hole dispersion. The real
part of $R({\bf k}\omega)$ can be calculated from the imaginary part
$\Im R({\bf k}\omega)$ and the Kramers-Kronig relation.

Notice that the interaction constant $g_{\bf kk'}$ is determined by the
Fourier transform of the hole hopping constant $t_{\bf nm}$. If the
hopping to the nearest and next nearest sites is taken into account the
constant acquires the form
\begin{equation}\label{intcon}
g_{\bf kk'}=t(\gamma_{\bf k'}-\gamma_{\bf k+k'})+t'(\gamma'_{\bf
k'}-\gamma'_{\bf k+k'}),
\end{equation}
where $\gamma'_{\bf k}=\cos(k_x)\cos(k_y)$. This constant vanishes for
${\bf k=Q}$ when the vector ${\bf k'}$ is located at the boundary of
the magnetic Brillouin zone. In other words, fermions near hot spots
interact weakly with spin excitations. This is connected with the
short-range character of the interaction described by constant
(\ref{intcon}) -- the decaying spin excitation on the site {\bf n}
creates the fermion pair on the same and neighbor sites which is
reflected in the above form of the interaction constant.

\begin{figure}[t]
\centerline{\includegraphics[width=6.5cm]{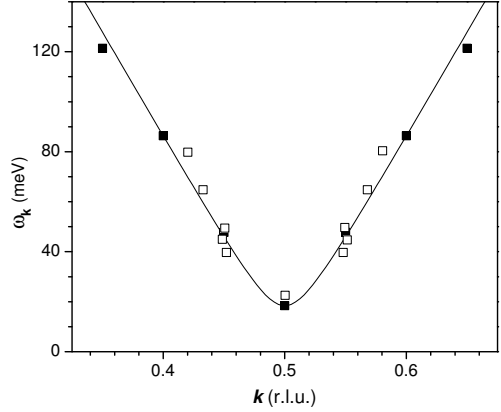}} \caption{The
dispersion of spin excitations calculated in a 20$\times$20 lattice for
$x=0.06$ and $T=17$~K (filled squares).\protect\cite{Sherman02} The
solid line is the fit of Eq.~(\protect\ref{mdisp}) to these data. Open
squares are the dispersion of the peak in the odd susceptibility in
YBa$_2$Cu$_3$O$_{6.5}$ ($x\approx 0.075$,
Ref.~\protect\refcite{Tallon}) at
$T=5$~K.\protect\cite{Bourges}}\label{Fig_i}
\end{figure}

As the quantity $\omega f_{\bf k}\Re R({\bf k}\omega)$ influences the
frequency of spin excitations only near {\bf Q}, it is convenient to
incorporate it in $\omega_{\bf k}$. This modifies the parameter
$\Delta>0$ which, as seen from Eqs.~(\ref{chi}) and (\ref{chitJ}),
describes a gap in the spin-excitation spectrum at the
antiferromagnetic wave vector {\bf Q}. The most exact way to determine
this parameter is to use the constraint of zero site magnetization
\begin{equation}\label{zsm}
\langle s^z_{\bf n}\rangle=\frac{1}{2}(1-x)-\langle s^{-1}_{\bf
n}s^{+1}_{\bf n}\rangle=0,
\end{equation}
which has to be fulfilled in the short-range antiferromagnetic
ordering. It can be shown that $\Delta\propto\xi^{-2}$ where $\xi$ is
the correlation length of the short-range order.\cite{Sherman03} Thus,
in this case the frequency of spin excitations at {\bf Q} is nonzero,
in contrast to the classical antiferromagnetic magnons. As follows from
Eq.~(\ref{chitJ}), the dispersion of spin excitations has a local
minimum at {\bf Q} and can be approximated as
\begin{equation}\label{mdisp}
\omega_{\bf k}=[\omega^2_{\bf Q}+c^2({\bf k-Q})^2]^{1/2}
\end{equation}
near this momentum. In Fig.~\ref{Fig_i} the calculated dispersion of
spin excitations\cite{Sherman02} near {\bf Q} is compared with the
dispersion of the maximum in the susceptibility in
YBa$_2$Cu$_3$O$_{6.5}$.\cite{Bourges} This is a bilayer crystal and the
symmetry allows one to divide the susceptibility into odd and even
parts. For the antiferromagnetic intrabilayer coupling the dispersion
of the maximum in the odd part can be compared with our calculations
carried out for a single layer. This comparison demonstrates that the
observed dispersion of the susceptibility maxima above $\omega_{\bf
Q}$, which we identify with the resonance frequency $\omega_r$, is
closely related to the dispersion of spin excitations.

Previous calculations \cite{Sherman03} show that the variation of the
temperature in the range from 0 to approximately 100~K leads only to
some broadening of maxima in the susceptibility. Therefore to simplify
calculations and use larger lattices, which is necessary to resolve the
low-frequency incommensurability, let us set $T=0$. In calculating $\Im
R({\bf k}\omega)$ the integration over frequencies in Eq.~(\ref{chitJ})
is the most time-consuming operation. For $T=0$ and $\omega\geq 0$ this
integral reduces to
$$\int_{-\omega}^0d\omega'\,A({\bf k'}\omega')
A({\bf k+k'},\omega+\omega')$$ and is easily integrated for the
spectral function (\ref{hsf}). The same result is obtained for
$\omega<0$, since $\Im R({\bf k}\omega)$ is an even function of
frequency. Notice that for $\eta\ll\omega$ the states with energies
\begin{equation}\label{energies}
-\omega<\varepsilon_{\bf k'}-\mu<0\;\;{\rm and}\;\; 0<\varepsilon_{\bf
k+k'}-\mu<\omega
\end{equation}
make the main contribution to this integral.

In the following, we use the values of $C_1$, $\Delta$ and $\alpha$
calculated self-con\-sis\-ten\-t\-ly in the $t$-$J$ model on a
20$\times$20 lattice for the range of hole concentrations $0\leq
x\lesssim 0.16$.\cite{Sherman03} The calculations were carried out for
the parameters $t=0.5$~eV and $J=0.1$~eV corresponding to hole-doped
cuprates.\cite{McMahan} In Eq.~(\ref{hsf}), for $\varepsilon_{\bf k}$
we apply the hole dispersion
\begin{eqnarray}
\varepsilon_{\bf k}&=&-0.0879+0.5547\gamma_{\bf k}-0.1327\gamma'_{\bf
 k}-0.0132\gamma_{2\bf k}\nonumber\\
&+&0.09245[\cos(2k_x)\cos(k_y)+\cos(k_x)\cos(2k_y)]
 -0.0265\gamma'_{2\bf k} \label{disp}
\end{eqnarray}
proposed from the analysis of photoemission data in
Bi$_2$Sr$_2$CaCu$_2$O$_8$.\cite{Norman} Here the coefficients are in
electronvolts. Results which are analogous to those discussed in the
next section can also be obtained with other model dispersions
suggested for cuprates.\cite{Liu,Brinckmann,Norman} Results do not
change qualitatively either with the variation of the parameter $t'$ in
Eq.~(\ref{intcon}) in the range from 0 to $-0.4t$ (notice that
parameters $t$ and $t'$ of the hole hopping part of Hamiltonian
(\ref{hamiltonian}) are only indirectly connected with the coefficients
in Eq.~(\ref{disp}), since to a great extent the hole dispersion is
shaped by the interaction between holes and spin
excitations\cite{Sherman04b}).

\section{Magnetic susceptibility}
\begin{figure}[t]
\centerline{\includegraphics[width=6.5cm]{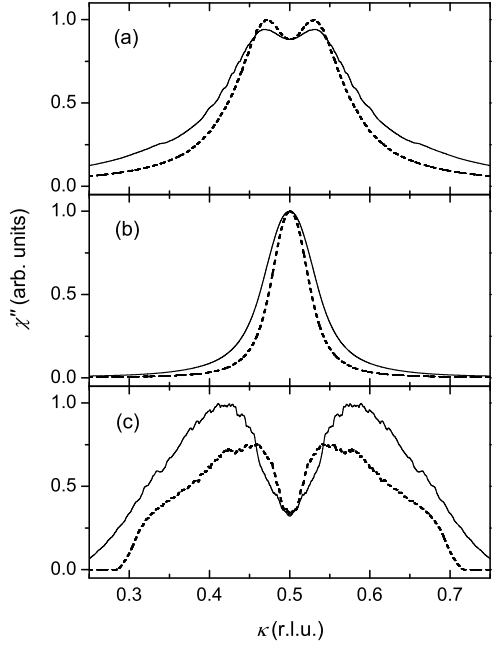}} \caption{The
momentum dependence of $\chi''({\bf k}\omega)$ for $T=0$, $x \approx
0.12$, $\mu=-40$~meV, $t'=-0.2t$ and $\omega=70$~meV, $\eta=30$~meV
(a), $\omega=35$~meV, $\eta=15$~meV (b), $\omega=2$~meV, $\eta=1.5$~meV
(c). Calculations were carried out in a 1200$\times$1200 lattice. The
solid lines correspond to the scans along the edge of the Brillouin
zone, ${\bf k}=(\kappa,\frac{1}{2})$; the dashed lines are for the zone
diagonal, ${\bf k}=(\kappa,\kappa)$.} \label{Fig_ii}
\end{figure}
\begin{figure}[t]
\centerline{\includegraphics[width=5cm]{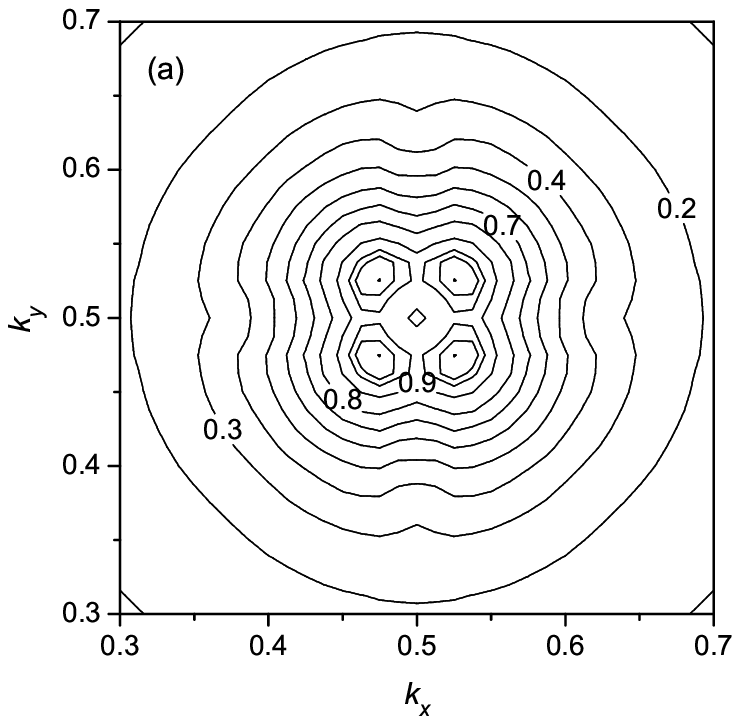}
\hspace{0.5cm}\includegraphics[width=5cm]{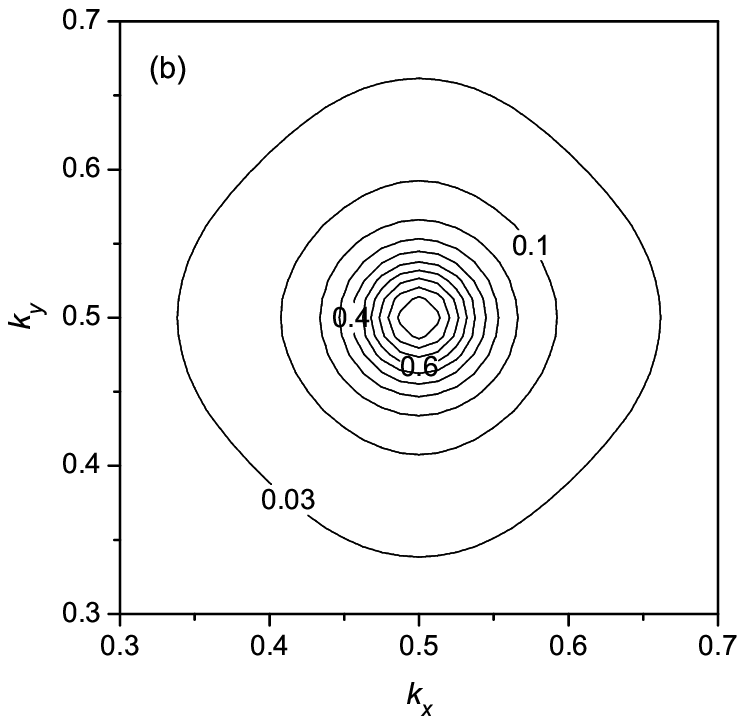}}
\vspace*{0.5cm}\centerline{\includegraphics[width=5cm]{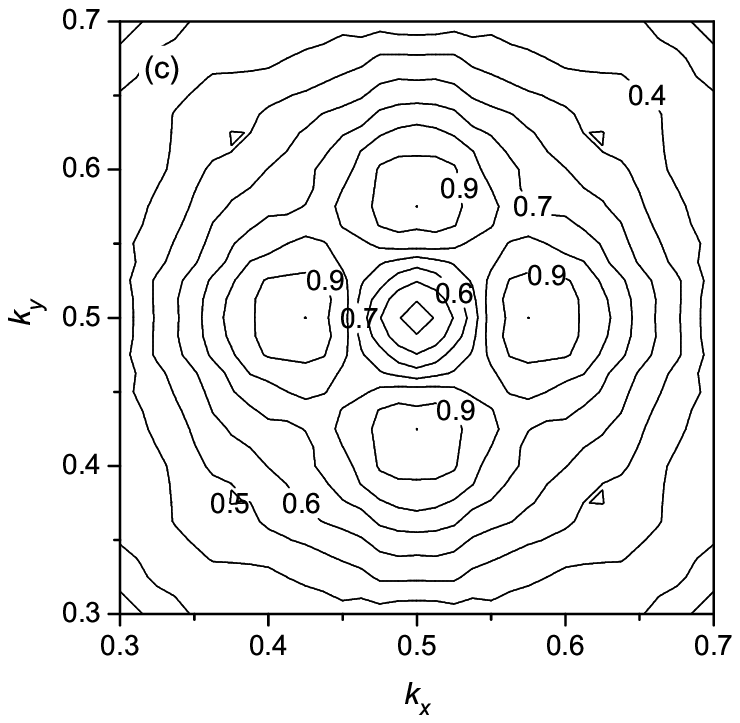}}
\caption{The contour plots of $\chi''({\bf k}\omega)$. Parameters in
parts (a), (b) and (c) are the same as in the respective parts of
Fig.~\protect\ref{Fig_ii}.} \label{Fig_iii}
\end{figure}
\begin{figure}[t]
\centerline{\includegraphics[width=6.5cm]{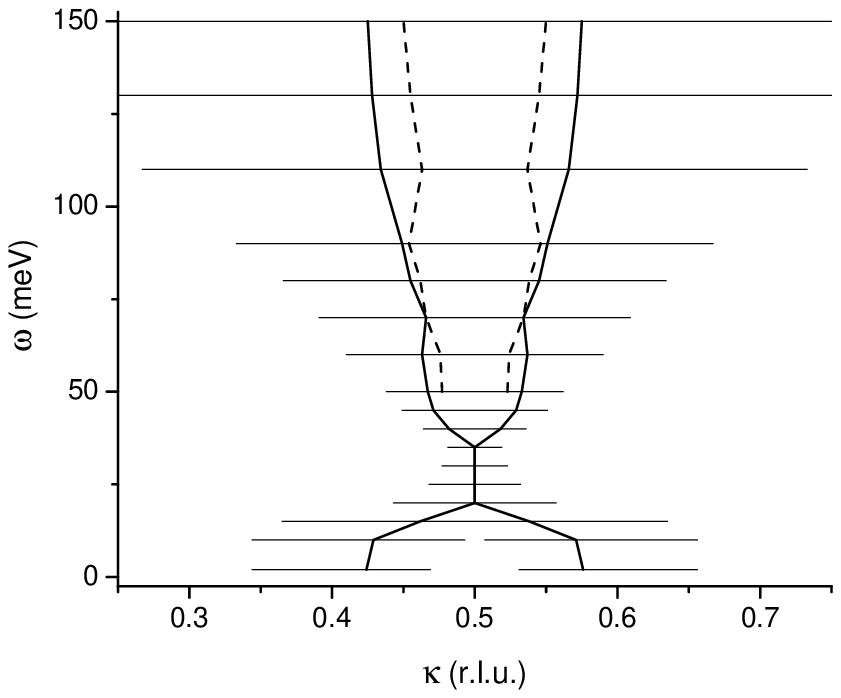}} \caption{The
dispersion of maxima in $\chi''({\bf k}\omega)$ for scans along the
edge [${\bf k}=(\kappa,\frac{1}{2})$, solid lines] and the diagonal
[${\bf k}=(\kappa,\kappa)$, dashed lines] of the Brillouin zone. The
dispersion along the diagonal is shown only in the frequency range in
which these maxima are more intensive than those along the edge.
Parameters are the same as in Fig.~\protect\ref{Fig_ii}. Horizontal
bars are FWHM for maxima along the edge of the Brillouin zone.}
\label{Fig_iv}
\end{figure}

The momentum dependence of $\chi''({\bf k}\omega)$ calculated with the
above equations for three energy transfers are shown in
Fig.~\ref{Fig_ii}. The contour plots of the susceptibility for the same
parameters are demonstrated in Fig.~\ref{Fig_iii}.
As seen from these figures, there are three frequency regions with
different shapes of the momentum dependence of $\chi''({\bf k}\omega)$.
The first region is the vicinity of the frequency $\omega_{\bf Q}$ of
the gap in the dispersion of spin excitations at the antiferromagnetic
wave vector {\bf Q}. For the parameters of Fig.~\ref{Fig_ii}
$\omega_{\bf Q}\approx 37$~meV. In this region the susceptibility is
peaked at the wave vector {\bf Q}. For smaller and larger frequencies
the magnetic response is incommensurate. The dispersion of maxima in
$\chi''({\bf k}\omega)$ for scans along the edge and the diagonal of
the Brillouin zone and their full widths at half maximum (FWHM) are
shown in Fig.~\ref{Fig_iv}. Analogous dispersion was obtained in
Ref.~\refcite{Norman} in the itinerant-carrier approach for the
superconducting state.

The momentum dependencies of the susceptibility which are similar to
those shown in Fig.~\ref{Fig_ii} and \ref{Fig_iii} were observed in
yttrium and lanthanum cuprates.\cite{Mason93,Dai,Tranquada} The
dispersion of the peaks in $\chi''({\bf k}\omega)$ which is similar to
that shown in Fig.~\ref{Fig_iv} was derived from experimental data in
YBa$_2$Cu$_3$O$_{7-y}$ and La$_{2-x}$Ba$_x$CuO$_4$ in
Refs.~\refcite{Dai,Tranquada}. As seen from Fig.~\ref{Fig_ii}, for
frequencies $\omega<\omega_{\bf Q}$ the susceptibility is peaked at the
wave vectors ${\bf k}=(\frac{1}{2},\frac{1}{2}\pm\delta)$ and
$(\frac{1}{2}\pm\delta,\frac{1}{2})$, while for $\omega>\omega_{\bf Q}$
the maxima are located at
$(\frac{1}{2}\pm\delta,\frac{1}{2}\pm\delta)$,
$(\frac{1}{2}\pm\delta,\frac{1}{2}\mp\delta)$ for the parameters used.
This result is also in agreement with experimental
observations.\cite{Dai,Hayden04,Tranquada} Notice, however, that for
$\omega>\omega_{\bf Q}$ the positions of maxima in the momentum space
may vary with parameters.

To understand the above results one should notice that Eq.~(\ref{chi})
contains the resonance denominator which will dominate in the momentum
dependence for $\omega \geq\omega_{\bf Q}$ if the spin excitations are
not overdamped. Parameters of Fig.~\ref{Fig_ii} correspond to this
case. For $\omega \geq\omega_{\bf Q}$ the equation $\omega=\omega_{\bf
k}$ determines the positions of the maxima in $\chi''({\bf k}\omega)$
which are somewhat shifted by the momentum dependence of the
spin-excitation damping $f_{\bf k}\Im R({\bf k}\omega)$. Using
Eq.~(\ref{mdisp}) we find that the maxima in $\chi''({\bf k}\omega)$
are positioned near a circle centered at {\bf Q} with the radius
$c^{-1}(\omega^2-\omega^2_{\bf Q})^{1/2}$.\cite{Barzykin,Sherman02}

In the region $\omega<\omega_{\bf Q}$ the nature of the
incommensurability is completely different. It is most easily seen in
the limit of small frequencies when Eq.~(\ref{chi}) reduces to
\begin{equation}\label{lowfreq}
\chi''({\bf k}\omega)\approx -4\mu_B^2\omega\frac{\Im R({\bf
k}\omega)}{\omega_{\bf k}^4}.
\end{equation}
As seen in Fig.~\ref{Fig_i}, $\omega^{-4}_{\bf k}$ is a decreasing
function of the difference ${\bf k-Q}$ which acts in favor of a
commensurate peak. However, if $\Im R({\bf k}\omega)$ in the numerator
of Eq.~(\ref{lowfreq}) has a pronounced dip at {\bf Q} the commensurate
peak splits into several incommensurate maxima. For hole concentrations
$x\lesssim 0.06$, when the Fermi surface consists of four ellipses
centered at
$(\pm\frac{1}{4},\pm\frac{1}{4})$,\cite{Izyumov,Sherman04b,Damascelli}
$\Im R({\bf k}\omega)$ has a dip at {\bf Q} due to the nesting of the
ellipses with this wave vector.\cite{Sherman04} For larger $x$ the
mechanism of the dip formation is the following.
\begin{figure}[t]
\centerline{\includegraphics[width=5cm]{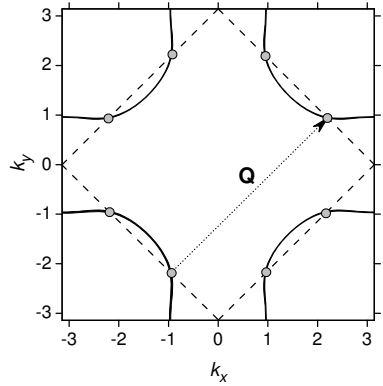}} \caption{The Fermi
surface for dispersion (\protect\ref{disp}) and $\mu=-40$~meV (solid
lines). Dashed lines show the boundary of the magnetic Brillouin zone,
gray circles are the hot spots, the dotted arrow is the
antiferromagnetic wave vector.} \label{Fig_v}
\end{figure}
As follows from Eq.~(\ref{energies}), for ${\bf k=Q}$ and small
frequencies $\omega$ hole states which make the main contribution to
the spin-excitation damping (\ref{chitJ}) are located near the hot
spots (see Fig.~\ref{Fig_v}). For these wave vectors the interaction
constant $g_{\bf Qk'}$, Eq.~(\ref{intcon}), is small which leads to the
smallness of $\Im R({\bf Q}\omega)$. With the wave vector moving away
from {\bf Q} momenta of states contributing to the spin-excitation
damping recede from the hot spots, the interaction constant grows, and
with it the spin-excitation damping. Thus, the damping has a dip at
{\bf Q} which leads to the low-frequency incommensurability shown in
Fig.~\ref{Fig_ii}c.

Let us compare the discussed mechanisms of the low- and high-frequency
incommensurability with those based on the picture of itinerant
electrons and the random phase approximation. In this latter approach
incommensurability arises due to maxima in the noninteracting
susceptibility $\chi_0$ described by the fermion
bubbles.\cite{Liu,Brinckmann,Norman} For low frequencies such a maximum
appears if the Fermi surface has nesting. As mentioned, this mechanism
imposes rather stringent requirements on the electron energy spectrum,
because to reproduce known experimental results the nesting has to
persist in the wide range of hole concentrations and the nesting
momentum has to change in a specific manner with doping. In
Ref.~\refcite{Brinckmann} the notion was proposed that the nesting for
constant-energy contours can appear in the superconducting state. The
application of this idea also requires fine tuning of
parameters.\cite{Norman} Besides, this mechanism cannot explain the
incommensurability above $T_c$ which is observed both in lanthanum and
yttrium cuprates.\cite{Mason93,Dai,Tranquada} In the approach discussed
in this paper requirements on the Fermi surface are substantially
relaxed: the Fermi surface has to intersect with the boundary of the
magnetic Brillouin zone, i.e. the Fermi surface has to contain hot
spots where the interaction constant $g_{\bf kk'}$ is small which leads
to the dip in ${\Im R}$ at {\bf Q}. According to the available
photoemission data\cite{Ino,Damascelli} Fermi surfaces of this type,
which resemble that shown in Fig.~\ref{Fig_v}, are indeed observed in
underdoped cuprates. Apart from Eq.~(\ref{disp}) we used some other
model dispersions present in the literature\cite{Liu,Brinckmann,Norman}
and obtained results which are qualitatively similar to those shown in
Fig.~\ref{Fig_ii} -- \ref{Fig_iv}. Hence the discussed mechanism is
robust with respect to changes of the hole energy spectrum, provided
that the Fermi surface contains hot spots. The mechanism is equally
applicable for the superconducting state, since the same interaction
constant $g_{\bf kk'}$ enters into the expression for the
susceptibility in this state.\cite{Sherman02} We suppose that in
certain conditions the magnetic incommensurability may trigger the
corrugation of Cu-O planes and formation of stripes.

\begin{figure}[t]
\centerline{\includegraphics[width=6.5cm]{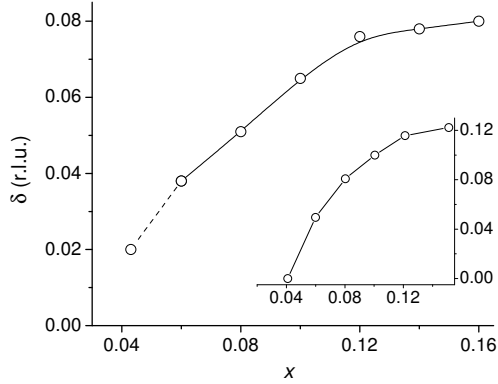}} \caption{The
incommensurability parameter $\delta$ vs.\ $x$ for $\omega=2$~meV. The
value of $\delta$ for $x=0.043$ was taken from
Ref.~\protect\refcite{Sherman04}. Inset: experimental
data\protect\cite{Yamada} for La$_{2-x}$Sr$_x$CuO$_4$. Connecting lines
are a guide to the eye.} \label{Fig_vi}
\end{figure}

The dependence of the incommensurability parameter $\delta$ on $x$ for
the low frequencies is shown in Fig.~\ref{Fig_vi}. In agreement with
experiment (see the inset in Fig.~\ref{Fig_vi}) $\delta$ grows nearly
linearly with $x$ up to $x\lesssim 0.12$ and then saturates. In this
calculation dispersion~(\ref{disp}) was used for the entire range of
hole concentrations $0.06\leq x\leq 0.16$. This is not quite correct,
since the photoemission data\cite{Damascelli} and
calculations\cite{Sherman04b} demonstrate that the dispersion changes
substantially with doping. However, we suppose that the growth of the
spin-excitation frequency with doping in accord with the relations
$\omega_{\bf Q}\propto\xi^{-1}\propto x^{1/2}$,\cite{Sherman03} which
leads to a weaker momentum dependence in Eq.~(\ref{mdisp}) and in the
denominator of Eq.~(\ref{lowfreq}), is more essential for the
dependence $\delta(x)$ than the variation of the hole dispersion.

We found also that the low-frequency incommensurability disappears when
the hole damping $\eta$ is greater than $\omega$. Besides, this
incommensurability disappears if the chemical potential $\mu$
approaches the extended van Hove singularities at $(0,\frac{1}{2})$,
$(\frac{1}{2},0)$. In this case for ${\bf k=Q}$ the entire region of
these singularities in which the interaction constant $g_{\bf Qk'}$ is
not small contributes to the spin-excitation damping. As a result the
dip in the damping becomes shallower or disappears completely. In the
$t$-$J$ model, $\mu$ approaches the van Hove singularities for
$x\approx 0.18$ for the parameters of hole-doped
cuprates.\cite{Sherman04b} This may be the reason of the disappearance
of the incommensurability in overdoped cuprates.\cite{Yamada} The
low-frequency incommensurability disappears also in lattices with size
less than 30$\times$30 sites. The use of a smaller lattice and an
increased artificial broadening needed for stabilizing the iteration
procedure accounts for the lack of low-frequency incommensurability in
the self-consistent calculations of Ref.~\refcite{Sherman03}.

\begin{figure}[t]
\centerline{\includegraphics[width=6.5cm]{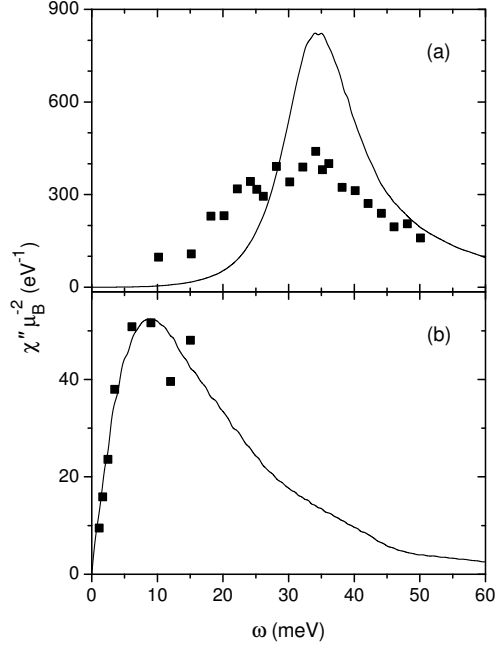}} \caption{The
frequency dependence of $\chi''$. The solid lines are our results for
$T=0$, $x \approx 0.12$, $\mu=-40$~meV, $\eta=3.5$~meV, ${\bf k=Q}$ (a)
and for ${\bf k}=(0.42,0.5)$ with the hole dispersion scaled by the
factor 0.4 (b, see text). Squares are the odd susceptibility
measured\protect\cite{Bourges} in the normal-state
YBa$_2$Cu$_3$O$_{6.83}$ ($x\approx 0.14$, \protect\refcite{Tallon}) at
$T=100$~K and ${\bf k=Q}$ (a) and the susceptibility in
La$_{1.86}$Sr$_{0.14}$CuO$_4$ for $T=35$~K at the incommensurate peak
\protect\cite{Aeppli} (b).} \label{Fig_vii}
\end{figure}
As mentioned above, for the parameters chosen spin excitations are not
overdamped near the antiferromagnetic wave vector. As a consequence,
the frequency dependence of $\chi''({\bf Q}\omega)$ has a pronounced
maximum at $\omega\approx\omega_{\bf Q}$ which resembles the
susceptibility observed in the underdoped YBa$_2$Cu$_3$O$_{7-y}$ in the
superconducting and normal states. As seen in Fig.~\ref{Fig_vii}a, the
experimental width of the maximum is approximately twice as large as
the calculated one. Partly this is connected with the difference in
temperatures for the two sets of the data. Besides, the width and shape
of the frequency dependence of $\chi''$ vary essentially with the
change of the hole dispersion and damping. Figure~\ref{Fig_vii}b
demonstrates that, in particular, the decrease of the hole bandwidth
leads to a substantial growth of the spin-excitation damping which in
its turn results in the overdamping of spin excitations. For this
figure the calculated results were obtained with dispersion
(\ref{disp}) scaled by the factor 0.4. The calculations were carried
out for ${\bf k}=(0.42,0.5)$ which corresponds to the wave vector of
the low-frequency peak in Fig.~\ref{Fig_ii}c. The overdamping of spin
excitations leads to the red shift of the maximum in $\chi''(\omega)$.
Its position is no longer connected with the frequency of spin
excitations. The similar frequency dependence of $\chi''$ without a
well-defined peak of spin excitations is observed in
La$_{2-x}$Sr$_x$CuO$_4$.\cite{Aeppli} Thus, we suppose that the
observed dissimilarity of the frequency dependencies of the
susceptibility in lanthanum and yttrium cuprates is connected with the
different values of the spin-excitation damping.

The increased spin-excitation damping obtained above with the scaled
hole dispersion does not affect markedly the low-frequency
incommensurability, however, for the frequencies $\omega_{\bf
Q}\geq\omega\geq 150$~meV we found only broad commensurate maxima
instead of the incommensurate peaks shown in Fig.~\ref{Fig_ii}a. Such
spectra are also observed experimentally.\cite{Endoh}

Our consideration was restricted to the normal state. In the considered
approach the opening of the superconducting gap suppresses the
spin-excitation damping for frequencies below the gap and increases the
damping above it. The respective redistribution of the intensity takes
place also in the
susceptibility.\cite{Mason93,Aeppli,Barzykin,Sherman02} As mentioned,
for the momentum dependence the same mechanisms which lead to the
incommensurate magnetic response in the normal state operate also in
the superconducting state. In this state the suppressed spin-excitation
damping produces sharper peaks in the susceptibility, however, their
location in the momentum space is approximately the same as in the
normal state.\cite{Dai,Kimura}

\section{Concluding remarks}
Mori's projection operator formalism and the $t$-$J$ model of Cu-O
planes were used for the interpretation of the magnetic susceptibility
in normal-state cu\-p\-ra\-te perovskites. It was shown that the
calculated momentum and frequency dependencies of the imaginary part of
the susceptibility $\chi''$, the dispersion and location of maxima in
it and the concentration dependence of the incommensurability parameter
are similar to those observed in lanthanum and yttrium cuprates. The
dispersion of the maxima in $\chi''$ resembles two parabolas with
upward- and downward-directed branches which converge at the
antiferromagnetic wave vector {\bf Q} and at the respective frequency
of spin excitations $\omega_{\bf Q}$. This frequency corresponds to a
local minimum in the dispersion of spin excitations and its value is
connected with the correlation length of the short-range
antiferromagnetic order. We relate the upper parabola to the
spin-excitation dispersion. The incommensurability connected with the
lower parabola is related to the dip in the spin-excitation damping at
{\bf Q}. For moderate doping the dip arises due to the smallness of the
interaction between spin excitations and holes near the hot spots,
which is a consequence of the short-range character of this
interaction. In agreement with experiment the incommensurate peaks
which form the lower parabola are located at momenta
$(\frac{1}{2},\frac{1}{2}\pm\delta)$ and
$(\frac{1}{2}\pm\delta,\frac{1}{2})$, while peaks in the upper parabola
are at $(\frac{1}{2}\pm\delta,\frac{1}{2}\pm\delta)$ and
$(\frac{1}{2}\pm\delta,\frac{1}{2}\mp\delta)$. Also in agreement with
experiment the low-frequency incommensurability parameter $\delta$
grows linearly with the hole concentration $x$ for $x\lesssim 0.12$ and
then saturates. This behavior of $\delta$ is mainly connected with the
concentration dependence of the frequency $\omega_{\bf Q}$ of the spin
gap at the antiferromagnetic wave vector. We found that the
incommensurability for the transfer frequencies $\omega<\omega_{\bf Q}$
disappears if the damping of holes with energies $\pm\omega$ is greater
than $\omega$. This incommensurability vanishes also when the chemical
potential approaches the extended van Hove singularities at
$(0,\frac{1}{2})$ and $(\frac{1}{2},0)$. The incommensurability for
$\omega>\omega_{\bf Q}$ disappears for large spin-excitation damping.
The value of this damping depends heavily on the hole damping and on
the shape and width of the hole band. We suppose that the marked
difference in the frequency dependencies of the susceptibility in
YBa$_2$Cu$_3$O$_{7-y}$ and La$_{2-x}$Sr$_x$CuO$_4$ -- a pronounced peak
at $\omega\approx 25-40$~meV for ${\bf k=Q}$ in the former crystal and
a broad feature at $\omega\approx 10$~meV in the latter -- is a
consequence of the difference in the electron spectra. The larger
spin-excitation damping in La$_{2-x}$Sr$_x$CuO$_4$ leads to overdamping
of spin excitations, while in the underdoped YBa$_2$Cu$_3$O$_{7-y}$ the
excitations are well-defined even in the normal state.

\section*{Acknowledgements}
This work was partially supported by the ESF grant No.~5548 and by the
DFG.

\end{document}